\documentclass{cupconf}
\input epsf

  \checkfont{eurm10}
  \iffontfound
    \IfFileExists{upmath.sty}
      {\typeout{^^JFound AMS Euler Roman fonts on the system,
                   using the 'upmath' package.^^J}%
       \usepackage{upmath}}
      {\typeout{^^JFound AMS Euler Roman fonts on the system, but you
                   dont seem to have the}%
       \typeout{'upmath' package installed. cupconf.cls can take advantage
                 of these fonts,^^Jif you use 'upmath' package.^^J}%
      }
  \else
  \fi

  \checkfont{msam10}
  \iffontfound
    \IfFileExists{amssymb.sty}
      {\typeout{^^JFound AMS Symbol fonts on the system, using the
                'amssymb' package.^^J}%
       \usepackage{amssymb}%
         \let\leq=\leqslant
         
      }{}
  \fi

  \IfFileExists{amsbsy.sty}
    {\typeout{^^JFound the 'amsbsy' package on the system, using it.^^J}%
     \usepackage{amsbsy}}
    {}

\newsavebox{\astrutbox}
\sbox{\astrutbox}{\rule[-5pt]{0pt}{20pt}}

\newcommand\etal{\mbox{\textit{et al.}}}

\newcommand\eg{e.g.\ }
\def\ie{{i.e.},~}  
\def\4he{$^4$He}  
\def\3he{$^3$He}  
\def\7li{$^7$Li}  
\def\Yp{Y$_{\rm P}$~}  
\def\yd{$y_{\rm D}$~}  
\def\y3{$y_{3}$~}  
\def\hii{H\thinspace{$\scriptstyle{\rm II}$}~}  
\def\hi{H\thinspace{$\scriptstyle{\rm I}$}~}  
\def\di{D\thinspace{$\scriptstyle{\rm I}$}~}  
\newcommand\la{\lower0.6ex\vbox{\hbox{\ensuremath{\buildrel{\textstyle<}\over{\sim}\  
}}}}  
\newcommand\ga{\lower0.6ex\vbox{\hbox{\ensuremath{\buildrel{\textstyle>}\over{\sim}\  
}}}}  
  
\newcommand{\obh}{\ensuremath{\Omega_{\rm B} h^2\;}}

\newcommand{\omb}{\ensuremath{\Omega_{\rm B}\;}}

\newcommand{\be}{\begin{equation}}  
\newcommand{\ee}{\end{equation}}  
\newcommand{\Deln}{\ensuremath{\Delta N_\nu\;}}  
\def\Nnu{$N_{\nu}$~}  
\newcommand{\epm}{\ensuremath{e^{\pm}\;}}  
  
\def\be{\begin{equation}}
\def\ee{\end{equation}}

\title[Primordial Nucleosynthesis]{Primordial Nucleosynthesis}

\author[G. Steigman]%
{G\ls A\ls R\ls Y\ns S\ls T\ls E\ls I\ls G\ls M\ls A\ls N}%

\affiliation{Department of Physics, The Ohio State University,
Columbus, OH 43210, USA\\}

\begin{document}

\maketitle

\begin{abstract}
The primordial abundances of deuterium, helium-3, helium-4, and lithium-7 
probe the baryon density of the Universe only a few minutes after the Big 
Bang.  Of these relics from the early Universe, deuterium is the baryometer 
of choice.  After reviewing the current observational status of the relic 
abundances (a moving target!), the baryon density determined by big bang 
nucleosynthesis (BBN) is derived.  The temperature fluctuation spectrum of 
the cosmic background radiation (CBR), established several hundred thousand 
years later, probes the baryon density at a completely different epoch in 
the evolution of the Universe.  The excellent agreement between the BBN- 
and CBR-determined baryon densities provides impressive confirmation of 
the standard model of cosmology, permitting the study of extensions of the 
standard model.  In combination with the BBN- and/or CBR-determined baryon 
density, the relic abundance of \4he provides an excellent chronometer, 
constraining those extensions of the standard model which lead to a 
nonstandard early-Universe expansion rate.
\end{abstract}

\firstsection 

\section{Introduction}

As the hot, dense, early Universe rushed to expand and cool, it briefly
passed through the epoch of big bang nucleosynthesis (BBN), leaving behind
as relics the first complex nuclei: deuterium, helium-3, helium-4, and
lithium-7.  The abundances of these relic nuclides were determined by the 
competition between the relative densities of nucleons (baryons) and photons 
and, by the universal expansion rate.  In particular, while deuterium is 
an excellent baryometer, \4he provides an accurate chronometer.  Nearly 
400 thousand years later, when the cosmic background radiation (CBR) had 
cooled sufficiently to allow neutral atoms to form, releasing the CBR from 
the embrace of the ionized plasma of protons and electrons, the spectrum 
of temperature fluctuations imprinted on the CBR encoded the baryon and 
radiation densities, along with the universal expansion rate at that epoch.  
As a result, the relic abundances of the light nuclides and the CBR 
temperature fluctuation spectrum provide invaluable windows on the early 
evolution of the Universe along with sensitive probes of its particle content.

The fruitful interplay between theory and data has been key to the enormous
progress in cosmology in recent times.  As new, more precise data became
available, models have had to be refined or rejected.  It is anticipated
this this process will -- indeed, should -- continue.  Therefore, this
review of the baryon content of the Universe as revealed by BBN and the 
CBR is but a signpost on the road to a more complete understanding of the 
history and evolution of the Universe.  By highlighting the current successes 
of the present ``standard'' model along with some of the challenges to it, 
I hope to identify those areas of theoretical and observational work which 
will contribute to continuing progress in our endeavor to understand the 
Universe, its past, present, and future.

\section{A BBN Primer}\label{sec:bbn}

Discussion of BBN can begin when the Universe is already a few tenths of 
a second old and the temperature is a few MeV.  At such early epochs the
Universe is too hot and dense to permit the presence of complex nuclei in 
any significant abundances and the baryons (nucleons) are either neutrons
or protons whose relative abundances are determined by the weak interactions
\be
p + e^{-} \longleftrightarrow n + \nu_{e}, ~n + e^{+} 
\longleftrightarrow p + \bar{\nu}_{e}, ~n \longleftrightarrow p + e^{-} 
+ \bar{\nu}_{e}.
\label{weak}
\ee
The higher neutron mass favors protons relative to neutrons, ensuring 
proton dominance.  When the weak interaction rates (Eq.~\ref{weak}) 
are fast compared to the universal expansion rate (and in the absence 
of a significant chemical potential for the electron neutrinos), $n/p 
\approx$ exp$(-\Delta m/T)$, where $\Delta m$ is the neutron-proton mass 
difference and $T$ is the temperature ($T_{\gamma} = T_{e} = T_{\nu} = 
T_{\rm N}$ prior to \epm annihilation).  If there were an {\it asymmetry} 
between the number densities of $\nu_{e}$ and $\bar{\nu}_{e}$ (``neutrino 
degeneracy''), described by a chemical potential $\mu_{e}$ (or, equivalently, 
by the dimensionless degeneracy parameter $\xi_{e} \equiv \mu_{e}/T$) then, 
early on, $n/p \approx$ exp$(-\Delta m/T -\xi_{e})$.  For a {\it significant} 
positive chemical potential ($\xi_{e} ~\ga 0.01$; more $\nu_{e}$ than 
$\bar{\nu}_{e}$) there are fewer neutrons than for the ``standard'' case 
(SBBN) which, as described below,  leads to the formation of less \4he.
 
The first step in building complex nuclei is the formation of deuterons
via $n + p \longleftrightarrow $D$ + \gamma$.  Sufficiently early on, when 
the Universe is very hot ($T ~\ga 80$~keV), the newly-formed deuterons find 
themselves bathed in a background of gamma rays (the photons whose relics 
have cooled today to form the CBR at a temperature of 2.7~K) and are quickly 
photo-dissociated, removing the platform necessary for building heavier 
nuclides. Only below $\sim 80$~keV has the Universe cooled sufficiently
to permit BBN to begin, leading to the synthesis of the lightest nuclides
D, \3he, \4he, and \7li. Once BBN begins, D, $^3$H, and \3he are rapidly
burned (for the baryon densities of interest) to \4he, the light nuclide
with the largest binding energy.  The absence of a stable mass-5 nuclide,
in combination with Coulomb barriers, suppresses the BBN production of
heavier nuclides; only \7li is synthesized in an astrophysically interesting
abundance.  All the while the Universe is expanding and cooling.  When
the temperature has dropped below $\sim 30$~keV, at a time comparable
to the neutron lifetime, the thermal energies of the colliding nuclides
is insufficient to overcome the Coulomb barriers, the remaining free
neutrons decay, and BBN ends.

From this brief overview of BBN it is clear that the relic abundances 
of the nuclides produced during BBN depend on the competition between 
the nuclear and weak interaction rates (which depend on the baryon 
density) and the universal expansion rate (quantified by the Hubble 
parameter $H$), so that the relic abundances provide early-Universe 
baryometers and chronometers.

\subsection{Early-Universe Expansion Rate}

The Friedman equation relates the expansion rate (measured by the Hubble 
parameter $H$) to the energy density ($\rho$): $H^{2} = {8\pi G \over 
3}\rho$ where, during the early, ``radiation-dominated'' (RD) evolution 
the energy density is dominated by the relativistic particles present 
($\rho = \rho_{\rm R}$).  For SBBN, prior to \epm annihilation, these 
are: photons, \epm pairs and, three flavors of left-handed (\ie one 
helicity state) neutrinos (and their right-handed, antineutrinos).
\be
\rho_{\rm R} = \rho_{\gamma} + \rho_{e} + 3\rho_{\nu} 
= {43 \over 8}\rho_{\gamma},
\label{rhopre}
\ee
where $\rho_{\gamma}$ is the energy density in CBR photons.  At this 
early epoch, when $T ~\la$ few MeV, the neutrinos are beginning to 
decouple from the $\gamma$ -- \epm plasma and the neutron to proton 
ratio, crucial for the production of primordial \4he, is decreasing.  
The time-temperature relation follows from the Friedman equation and 
the temperature dependence of $\rho_{\gamma}$
\be
{\rm Pre-\epm annihilation}:~~t~T_{\gamma}^{2} = 0.738~{\rm MeV^{2}~s}.
\label{ttpre}
\ee
To a very good (but not exact) approximation the neutrinos ($\nu_{e}$, 
$\nu_{\mu}$, $\nu_{\tau}$) are decoupled when the \epm pairs annihilate 
as the Universe cools below $m_{e}c^{2}$.  In this approximation the 
neutrinos don't share in the energy transferred from the annihilating 
\epm pairs to the CBR photons so that in the post-\epm annihilation 
universe the photons are hotter than the neutrinos by a factor 
$T_{\gamma}/T_{\nu} = (11/4)^{1/3}$, and the relativistic energy 
density is
\be
\rho_{\rm R} = \rho_{\gamma} + 3\rho_{\nu} = 1.68\rho_{\gamma}.
\label{rhopost}
\ee
The post-\epm annihilation time-temperature relation is
\be
{\rm Post-\epm annihilation}:~~t~T_{\gamma}^{2} = 1.32~{\rm MeV^{2}~s}.
\label{ttpost}
\ee

\subsubsection{~Additional Relativistic Energy Density}\label{sec:sneq1}

One of the most straightforward variations of the standard model of cosmology 
is to allow for an early (RD) nonstandard expansion rate $H' \equiv SH$,
where $S \equiv H'/H = t/t'$ is the {\it expansion rate factor}.  One
possibility for $S \neq 1$ is from the modification of the RD energy density
(see Eqs.~\ref{rhopre} \& \ref{rhopost}) due to ``extra'' relativistic 
particles $X$: $\rho_{\rm R} \rightarrow \rho_{\rm R} + \rho_{X}$.  If the 
extra energy density is normalized to that which would be contributed by 
one additional flavor of (decoupled) neutrinos (Steigman, Schramm \& Gunn 
1977), $\rho_{X} \equiv \Delta$N$_{\nu} \rho_{\nu}$ (N$_{\nu} \equiv 3 + 
\Delta$N$_{\nu}$), then
\be
S_{pre} \equiv (H'/H)_{pre} = (1+0.163\Delta N_{\nu})^{1/2}\,; \, \,\, 
S_{post} \equiv (H'/H)_{post} = (1+0.135\Delta N_{\nu})^{1/2}.
\label{sx}
\ee
Notice that $S$ and \Deln are related {\it nonlinearly}.  It must be 
emphasized that it is $S$ and not \Deln that is the fundamental parameter 
in the sense that {\it any} term in the Friedman equation which scales as 
radiation, decreasing with the fourth power of the scale factor, will 
change the standard-model expansion rate ($S \neq 1$).  For example, 
higher-dimensional effects such as in the Randall-Sundrum model (Randall 
\& Sundrum 1999a) may lead to either a speed-up in the expansion rate 
($S > 1$; $\Delta$N$_{\nu} > 0$) or, to a slow-down ($S < 1$; 
$\Delta$N$_{\nu} < 0$); see, also, \cite{rsb}, \cite{bin}, \cite{cline}.

\subsection{The Baryon Density}

In the expanding Universe, the number densities of all particles decrease 
with time, so that the magnitude of the baryon density (or that of any other 
particle) has no meaning without also specifying {\it when it is measured}.  
To quantify the universal abundance of baryons, it is best to compare 
$n_{\rm B}$ to the CBR photon density $n_{\gamma}$.  The ratio, $\eta 
\equiv n_{\rm B}/n_{\gamma}$ is very small, so that it is convenient to 
define a quantity of order unity,
\be
\eta_{10} \equiv 10^{10}(n_{\rm B}/n_{\gamma}) = 274\,\Omega_{\rm B}h^{2}
\equiv 274\,\omega_{\rm B},
\label{eq:eta10}
\ee
where \omb is the ratio (at present) of the baryon density to the critical
density and $h$ is the present value of the Hubble parameter in units of
100 km s$^{-1}~$Mpc$^{-1}$ ($\omega_{\rm B} \equiv \Omega_{\rm B}h^{2}$).  

\section{BBN Abundances}\label{sec:abund}

\begin{figure}
\centering
 \epsfysize=4.0truein
  \epsfbox{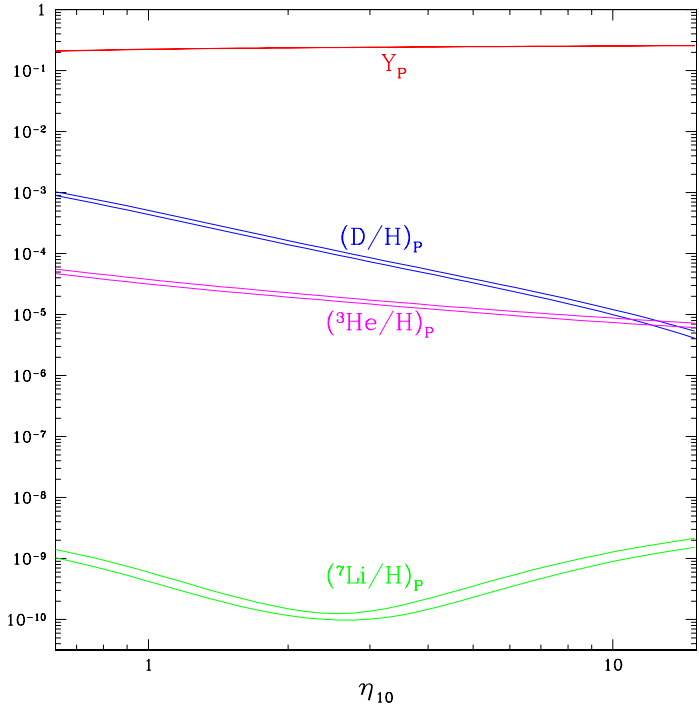}
\caption{The SBBN-predicted abundances of D, \3he, and \7li by number
with respect to hydrogen, and the \4he mass fraction Y$_{\rm P}$, as 
a function of the nucleon (baryon) abundance parameter $\eta_{10}$.  
The bands reflect the theoretical uncertainties ($\pm 1\sigma$) in 
the BBN predictions.
\label{fig:schrplot}}
\end{figure}

The relic abundances of D, \3he, and \7li are {\it rate limited}, determined 
by the competition between the early Universe expansion rate and the nucleon 
density.  Any of these three nuclides is, therefore, a potential baryometer; 
see Figure~\ref{fig:schrplot}.  

In contrast to the synthesis of the other light nuclides, once BBN begins 
($T ~\la 80$~keV) the reactions building \4he are so rapid that its relic 
abundance is not rate limited. The primordial abundance of \4he is limited 
by the availability of neutrons.  To a very good approximation, its relic 
abundance is set by the neutron abundance at the beginning of BBN.  As a 
result, the primordial mass fraction of \4he, Y$_{\rm P}$, while being 
a relatively insensitive baryometer (see Figure~\ref{fig:schrplot}), is 
an excellent, early-Universe chronometer.  

The qualitative effects of a nonstandard expansion rate on the relic
abundances of the light nuclides may be understood with reference to
Figure~\ref{fig:schrplot}.  For the baryon abundance range of interest 
the relic abundances of D and \3he are {\it decreasing} functions of 
$\eta$; in this range, D and \3he are being destroyed to build \4he.  
A faster than standard expansion ($S > 1$) provides less time for this 
destruction so that more D and \3he will survive.  The same behavior 
occurs for \7li at low values of $\eta$, where its abundance is a 
decreasing function of $\eta$.  However, at higher values of $\eta$, 
the BBN-predicted \7li abundance {\it increases} with $\eta$, so that 
less time available results in less production and a {\it smaller} \7li 
relic abundance.  Except for dramatic changes to the early-Universe 
expansion rate, these effects on the relic abundances of D, \3he, and 
\7li are subdominant to their variations with the baryon density.  Not 
so for \4he, whose relic abundance is very weakly (logarithmically) 
dependent on the baryon density, but very strongly dependent on the 
early-Universe expansion rate.  A faster expansion leaves more neutrons 
available to build \4he; to a good approximation $\Delta$Y $\approx 
0.16\,(S-1)$.  It is clear then that if \4he is paired with any of 
the other light nuclides, together they can constrain the baryon 
density ($\eta$ or \obh $\equiv \omega_{\rm B}$) and the early-Universe 
expansion rate ($S$ or $\Delta$N$_{\nu}$).

As noted above in \S \ref{sec:bbn}, the neutron-proton ratio at BBN can 
also be modified from its standard value in the presence of a significant 
electron-neutrino asymmetry ($\xi_{e} ~\ga 0.01$).  As a result, \Yp is 
also sensitive to any neutrino asymmetry.  More $\nu_{e}$ than $\bar{
\nu}_{e}$ drives the neutron-to-proton ratio down (see Eq.~\ref{weak}), 
leaving fewer neutrons available to build \4he; to a good approximation 
$\Delta$Y $\approx -0.23\,\xi_{e}$ (Kneller \& Steigman 2003).  In contrast, 
the relic abundances of D, \3he, and \7li are very insensitive to $\xi_{e} 
\neq 0$, so that when paired with \4he, they can simultaneously constrain 
the baryon density and the electron-neutrino asymmetry.  Notice that if 
{\it both} $S$ and $\xi_{e}$ are allowed to be free parameters, another 
observational constraint is needed to simultaneously constrain $\eta$, 
$S$, and $\xi_{e}$.  While neither \3he nor \7li can provide the needed 
constraint, the CBR temperature anisotropy spectrum, which is sensitive 
to $\eta$ and $S$, but not to $\xi_{e}$, can (see Barger \etal~2003b).
This review will concentrate on combining constraints from the CBR and
SBBN ($S = 1$, $\xi_{e} = 0$) and also for $S \neq 1$ ($\xi_{e} = 0$).
For the influence of and constraints on electron neutrino asymmetry,
see \cite{barger2} and further references therein. 

\section{Relic Abundances}

BBN constraints on the universal density of baryons and on the early-Universe
expansion rate require reasonably accurate determinations of the relic
abundances of the light nuclides.  As already noted, D, \3he, and \7li 
are all potential baryometers, while \4he is an excellent chronometer.
The combination of the availablility of large telescopes and advances
in detector technology has made it possible to obtain abundance estimates
at various sites in the Galaxy and elsewhere in the Universe with 
unprecedented precision (statistically).  However, the path to accurate
primordial abundances is littered with systematic uncertainties which
have the potential to contaminate otherwise exquisite data.  It is,
therefore, fortunate that the relic nuclides follow very different 
post-BBN evolutionary paths and are observed in diverse environments 
using a wide variety of astronomical techniques.  Neutral deuterium is
observed in absorption in the UV (or, in the optical when redshifted)
against background, bright sources (O or B stars in the Galaxy, QSOs
extragalactically).  Singly-ionized helium-3 is observed in emission 
in Galactic \hii regions via its spin-flip transition (the analog of
the 21 cm line in neutral hydrogen).  The helium-4 abundance is largely
determined by observations of recombination lines of ionized (singly and
doubly) \4he compared to those of ionized hydrogen in Galactic and,
especially, extragalactic \hii regions.  Observations of \7li, at
least those at low metallicity (nearly primordial) are restricted to
absorption in the atmospheres of the oldest, most metal-poor stars in 
the halo of the Galaxy.  The different evolutionary histories (described
below) combined with the differrent observational strategies provide a
measure of insurance that systematic errors in the determination of one
of the light element abundances are unlikely to propagate into other 
abundance determinations.

\subsection{Deuterium -- The Baryometer Of Choice}\label{sec:bbnbaryometer}

The deuteron is the most weakly bound of the light nuclides.  As a result,
any deuterium cycled through stars is burned to \3he and beyond.  Thus,
its post-BBN evolution is straightforward: deuterium observed anywhere, 
anytime, provides a {\it lower} bound to the primordial D abundance.  
For ``young'' systems, in the sense of little stellar evolution (\eg 
sites at high redshift and/or with very low metallicity), the observed
D abundance should reach a plateau at the primordial value.  Although
there are observations of deuterium in the solar system and the interstellar
medium (ISM) of the Galaxy which provide interesting lower bounds to
its primordial abundance, the observations of relic D in a few, high 
redshift, low metallicity, QSO absorption line systems (QSOALS) are 
of most value in estimating its primordial abundance.  

\begin{figure}
\centering
 \epsfysize=4.0truein
  \epsfbox{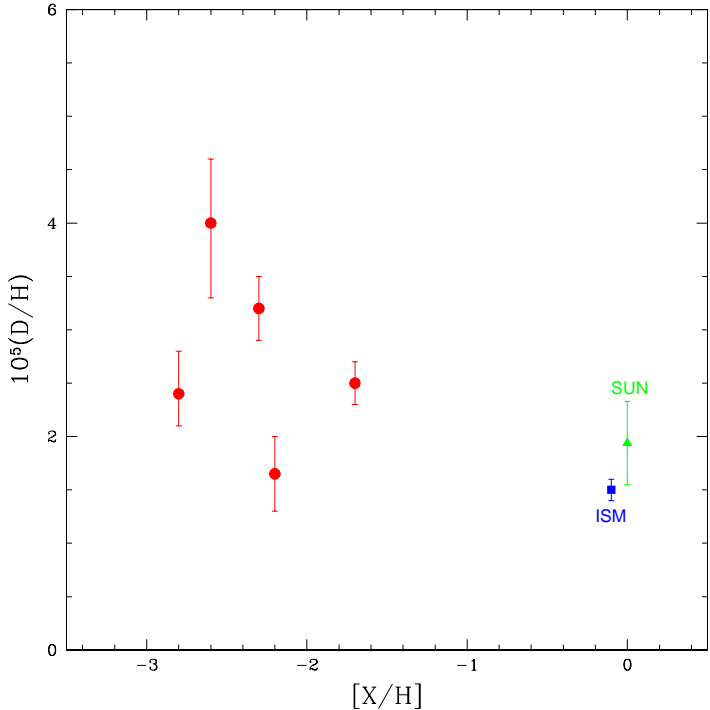}
\caption{Deuterium abundances, by number with respect to hydrogen 
D/H, versus metallicity (relative to solar on a log scale) from 
observations (as of early 2003) of QSOALS (filled circles).  
``X'' is usually silicon  or oxygen.  Shown for comparison are 
the D abundances inferred for the local ISM (filled square) and 
the solar system (presolar nebula: ``Sun''; filled triangle).    
\label{fig:dvssi}}  
\end{figure}    

While its simple post-BBN evolution is the greatest asset for relic 
D, the identical absorption spectra of \di and \hi (except for the 
velocity/wavelength shift resulting from the heavier reduced mass 
of the deuterium atom) is a severe liability, limiting significantly
the number of useful targets in the vast Lyman-alpha forest of the
QSO absorption spectra (see \cite{kirk} for a discussion).  It is
essential in choosing a target QSOALS that its velocity structure be
``simple'' since a low column density \hi absorber, shifted by $\sim
81$~km/s with respect to the main \hi absorber (an ``interloper'') 
would masquerade as \di absorption.  If this is not recognized, a too 
high D/H ratio would be inferred.  Since there are many more low-column 
density absorbers than those with high \hi column densities, absorption 
systems with somewhat lower \hi column density (\eg Lyman-limit systems: 
LLS) are more susceptible to this contamination than the higher \hi 
column density absorbers (\eg damped Ly$\alpha$ absorbers: DLA).  
However, while the DLA have many advantages over the LLS, a precise 
determination of the \hi column density requires an accurate placement 
of the continuum, which could be compromised by interlopers.  This might 
lead to an overestimate of the \hi column density and a concomitant 
underestimate of D/H (J. Linsky, private communication).  As a result
of these complications, the path to primordial D using  QSOALS  has 
not been straightforward, and some abundance claims have had to be 
withdrawn or revised.  Presently there are only five QSOALS with 
reasonably firm deuterium detections~\cite{kirk} (and references 
therein); these are shown in Figure~\ref{fig:dvssi} along with the 
corresponding solar system and ISM D abundances.  It is clear from 
Figure~\ref{fig:dvssi}, that there is significant dispersion among 
the derived D abundances at low metallicity which, so far, mask the 
anticipated deuterium plateau.  This suggests that systematic errors of 
the sort described here may have contaminated some of the determinations 
of the \di and/or \hi column densities. 

\begin{figure}
\centering
 \epsfysize=4.0truein
  \epsfbox{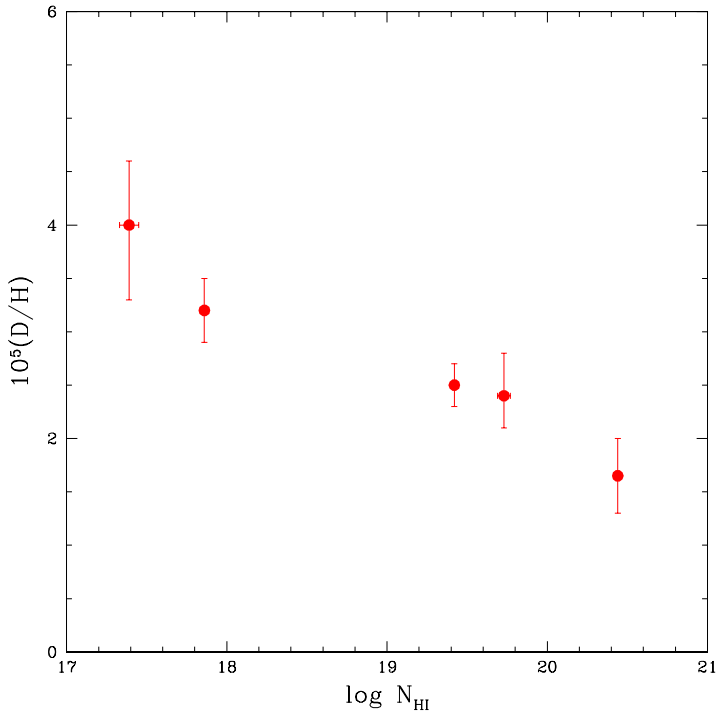}
\caption{Deuterium abundances versus the \hi column densities for 
the corresponding QSOALS shown in Figure \ref{fig:dvssi}.
\label{fig:dvsh}}  
\end{figure}    

To explore the possibility that such systematic effects, which would be 
correlated with the \hi column density, may be responsible for at least 
some of the dispersion revealed in Figure~\ref{fig:dvssi}, it is useful 
to plot the same QSOALS data versus the \hi column density;  this is shown 
in Figure \ref{fig:dvsh}.  Indeed, there is the suggestion from this very
limited data set that the low column density absorbers (LLS) have high
D/H, while the high column density systems (DLA) have low abundances.
However, on the basis of extant data it is impossible to decide which, 
if any, of these systems has been contaminated; there is no justification
for excluding any of the present data.  Indeed, perhaps the data is
telling us that our ideas about post-BBN deuterium evolution need to
be revised.

To proceed further using the current data I follow the lead of \cite{omear} 
and \cite{kirk} and adopt for the primordial D abundance the weighted mean 
of the D abundances for the five lines of sight (Kirkman \etal~2003); the 
dispersion in the data is used to set the error in $y_{\rm D}$: \yd = 
$2.6 \pm 0.4$.  It should be noted that using the same data \cite{kirk} 
derive a slightly higher mean D abundance: \yd = 2.74.  The difference 
is traced to their first finding the mean of log(y$_{\rm D}$) and then 
using it to compute the mean D abundance (\yd $ \equiv 10^{\langle 
\log(y_{\rm D})\rangle}$). 

The BBN-predicted relic abundance of deuterium depends sensitively on
the baryon density, \yd $\propto \eta^{-1.6}$, so that a ~$\sim 10\%$
determination of \yd can be used to estimate the baryon density to
$\sim 6\%$. For SBBN ($S = 1$ (\Nnu = 3), $\xi_{e} = 0$), the adopted
primordial D abundance corresponds to $\eta_{10}({\rm SBBN}) = 6.10^
{+0.67}_{-0.52}$ (\obh $= 0.0223^{+0.0024}_{-0.0019}$), in spectacular
agreement with the \cite{sperg} estimate of $\eta_{10} = 6.14 \pm 0.25$ 
(\obh $= 0.0224 \pm 0.0009$) based on WMAP and other CBR data (ACBAR 
and CBI) combined with large scale structure (2dFGRS) and Lyman-alpha 
forest constraints.  Indeed, if the \cite{sperg} estimate is used for
the BBN baryon density, the BBN-predicted deuterium abundance is \yd
$= 2.57 \pm 0.27$ (where a generous allowance of $\sim 8\%$ has been 
made for the uncertainty in the BBN prediction at fixed $\eta$; for 
the \cite{bnt} nuclear cross sections and uncertainties the result
is \yd $= 2.60^{+0.20}_{-0.18}$). 

\subsection{Helium-3}

Unlike D, the post-BBN evolution of \3he and \7li are quite complex.  
\3he is destroyed in the hotter interiors of all but the least massive 
(coolest) stars, but it is preserved in the cooler, outer layers of 
most stars.  In addition, hydrogen burning in low mass stars results 
in the production of significant amounts of {\it new} \3he (Iben 1967; 
Rood 1972; Dearborn, Steigman \& Schramm 1986; Vassiliadis \& Wood 1993; 
Dearborn, Steigman \& Tosi 1996).  To follow the post-BBN evolution of 
\3he, it is necessary to account for all these effects -- quantitatively 
-- in the material returned by stars to the interstellar medium (ISM).  
As indicated by the existing Galactic data (Geiss \& Gloeckler 1998; 
Bania, Rood \& Balser 2002), a very delicate balance exists between net 
production and net destruction of \3he in the course of the evolution of 
the Galaxy.  As a consequence, aside from noting an excellent qualitative 
agreement between the SBBN predicted and observed \3he abundances, \3he 
has -- at present -- little role to play  as a quantitatively useful 
baryometer. In this spirit, it is noted that an uncertain estimate of
the primordial abundance of \3he may be inferred from the observation
of an outer-Galaxy (less evolved) \hii region (Bania \etal~2002): \y3
$\equiv 10^{5}(^{3}$He/H) = $1.1 \pm 0.2$.

\subsection{Lithium-7}

A similar scenario may be sketched for \7li.  As a weakly bound nuclide, 
it is easily destroyed when cycled through stars except if it can be kept 
in the cooler, outer layers.  The high lithium abundances observed in 
the few ``super-lithium-rich red giants'' provide direct evidence that 
at least some stars can synthesize post-BBN lithium and bring it to the 
surface.  But, an unsolved issue is how much of this newly-synthesized 
lithium is actually returned to the ISM rather than mixed back into the
interior and destroyed.  

\begin{figure}
 \centering
  \vspace{-7.0pc}
   \epsfysize=4.283truein
    \epsfbox{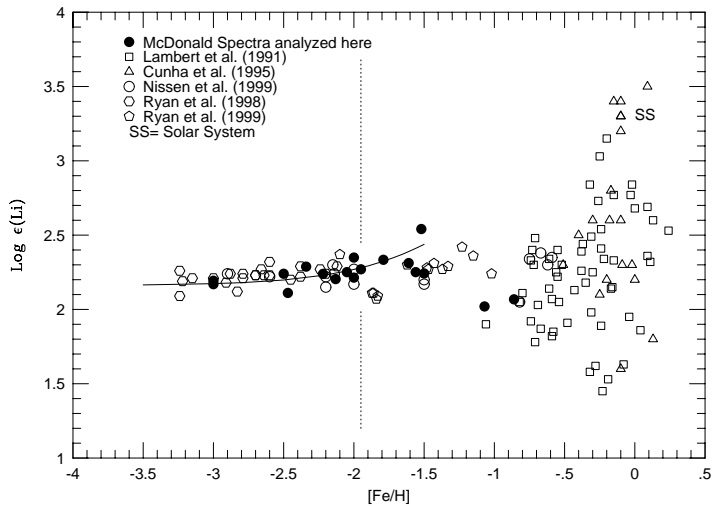}
\caption{Lithium abundances, log~$\epsilon$(Li)~$ \equiv$~[Li] $\equiv 
12 + $log(Li/H) versus metallicity (on a log scale relative to solar) 
from a compilation of stellar observations by V. V. Smith.
\label{fig:livsfe}}  
\end{figure}    

With these caveats in mind, in Figure~\ref{fig:livsfe} lithium abundances 
are shown as a function of metallicity from a compilation by V. V. Smith 
(private communication).  Since the quest for nearly primordial lithium is 
restricted to the oldest, most metal-poor stars in the Galaxy, stars that 
have had the most time to redistribute -- and destroy or dilute -- their 
surface lithium abundances, it is unclear whether the ``plateau'' at low
metallicities is representative of the primordial abundance of lithium.  
Although it seems clear that the lithium abundance in the Galaxy has 
increased since BBN, a quantitatively reliable estimate of its primordial 
abundance eludes us at present.  Given this state of affairs, the most 
fruitful approach is to learn about stellar structure and evolution by 
comparing the BBN-predicted lithium abundance to those abundances inferred 
from observations of the oldest stars, rather than to attempt to use 
the stellar observations to constrain the BBN-inferred baryon density.  
Concentrating on the low-metallicity, nearly primordial data, it seems 
that [Li] $\equiv 12 + $log(Li/H) $\approx 2.2 \pm 0.1$.  This estimate 
will be compared to the BBN-predicted lithium abundance using D as a 
baryometer and, to the BBN-predicted lithium abundance using the CBR-inferred 
baryon density.  Any tension between these BBN-predicted abundances and 
that inferred from the Galactic data may provide hints of nonstandard 
stellar astrophysics.

\subsection{Helium-4 -- The BBN Chronometer}

\begin{figure}
\centering
\vspace{1.5pc} 
 \epsfysize=3.25truein
  \epsfbox{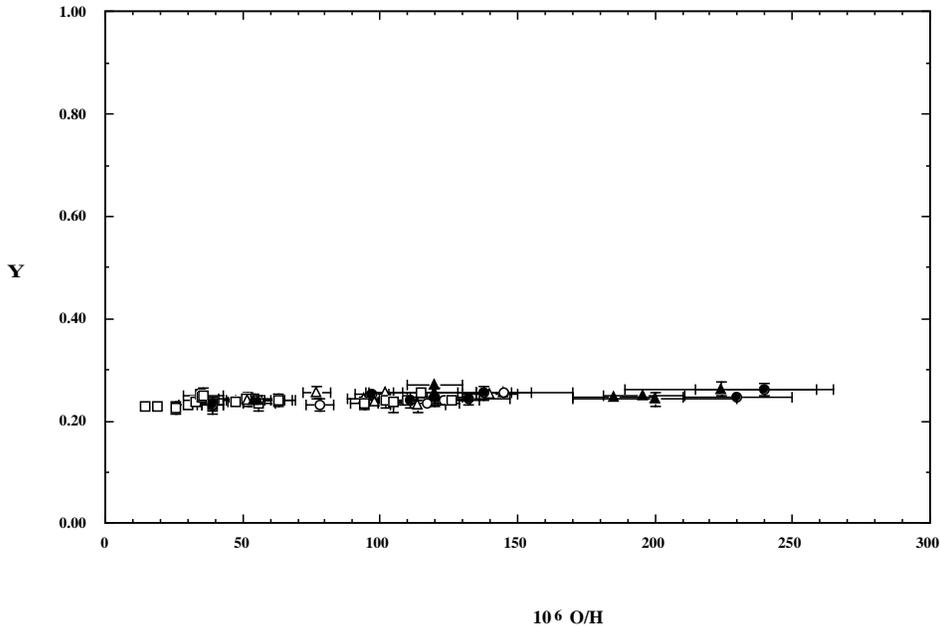}
\caption{The \4he mass fraction Y derived from observations 
of extragalactic \hii regions of low metallicity versus the 
corresponding \hii region oxygen abundances (from K. A. Olive).
\label{fig:yvso}}  
\end{figure}    

The good news about \4he is that, as the second most abundant nuclide, 
it may be observed throughout the Universe.  The bad news is that its 
abundance has evolved since the end of BBN.  In order to infer its 
primordial value it is therefore necessary to track the \4he abundance 
determinations (mass fraction Y$_{\rm P}$) as a function of metallicity 
or, to limit observations to very low metallicity objects.  Although, 
as for D, there are observations of \4he in the ISM and the solar system, 
the key data for determining its primordial abundance comes from observations 
of metal-poor, extragalactic \hii regions.  A compilation of current data 
(courtesy of K. A. Olive) is shown in Figure \ref{fig:yvso} where the 
\4he mass fraction is plotted as a function of the oxygen abundance; 
note that the solar oxygen abundance, O/H $\approx 5 \times 10^{-4}$ 
(Allende-Prieto, Lambert \& Asplund 2001) is off-scale in this figure. 
These are truly low metallicity \hii regions. 

It is clear from Figure \ref{fig:yvso} that the data exist to permit 
the derivation of a reasonably accurate estimate (statistically) of 
the primordial \4he mass fraction Y$_{\rm P}$, with or without  any 
extrapolation to zero-metallicity.  What is not easily seen in Figure 
\ref{fig:yvso} given the \Yp scale, is that \Yp derived from the data 
assembled from the literature by \cite{os} and \cite{oss} (\Yp = $0.234 
\pm 0.003$) is marginally inconsistent (at $\sim 2\sigma$) with the value 
derived by \cite{itl} and \cite{it} from their nearly independent data 
set (\Yp = $0.244 \pm 0.002$).  In addition, there are a variety of 
systematic corrections which might modify {\it both} data sets (Steigman, 
Viegas \& Gruenwald 1997; Viegas, Gruenwald \& Steigman 2000; Olive \& 
Skillman 2001; Sauer \& Jedamzik 2002; Gruenwald, Steigman \& Viegas 2002; 
Peimbert, Peimbert \& Luridiana 2002)

Unless/until the differences in \Yp derived by different authors from
somewhat different data sets is resolved and the known systematic errors 
are corrected for (the unknown ones will always hang over us like the 
sword of Damocles), the following compromise, adopted by \cite{osw}, may 
not be unreasonable.  From \cite{os} and \cite{oss}, the $2\sigma$ range 
for \Yp is 0.228 -- 0.240, while from the \cite{itl} and \cite{it} data 
the $2\sigma$ range is \Yp = 0.240 -- 0.248.  Thus, although the current
estimates are likely dominated by systematic errors, they span a $\sim 
2\sigma$ range from \Yp = 0.228 to \Yp = 0.248.  Therefore, as proposed 
by \cite{osw}, we adopt here a central value for \Yp = 0.238 and a $\sim 
1\sigma$ uncertainty of 0.005: \Yp = $0.238 \pm 0.005$.  Given the 
approximation (see \S \ref{sec:abund}) $\Delta$Y $\approx 0.16\,(S-1)$, 
for $\sigma_{\rm Y_{\rm P}} \approx 0.005$ the uncertainty in $S$ is 
$\approx 0.03$ (corresponding to an uncertainty in \Deln of $\approx 0.4$).

\section{The Baryon Density From SBBN}\label{sec:sbbn}

For SBBN, where $S = 1$ (\Nnu = 3) and $\xi_{e} = 0$, the primordial
abundances of D, \3he, \4he, and \7li are predicted as a function 
of only one free parameter, the baryon density parameter ($\eta$ 
or \obh $\equiv \omega_{\rm B}$).  As described above (see \S \ref
{sec:bbnbaryometer}), D is the baryometer of choice.  
From SBBN and the adopted relic abundance of deuterium, \yd $= 2.6 
\pm 0.4$, $\eta_{10} = 6.1^{+0.7}_{-0.5}$ (\obh $= 0.022 \pm 0.002$).


Having determined the baryon density to $\sim 10\%$ using D as the SBBN 
baryometer, it is incumbent upon us to compare the SBBN-predicted abundances 
of the other light nuclides with their relic abundances inferred from the 
observational data.  For this baryon density, the predicted primordial 
abundance of \3he is $y_{3} = 1.04 \pm 0.10$, in excellent agreement with 
the primordial value of $y_{3} = 1.1 \pm 0.2$ inferred from observations 
of an outer-Galaxy \hii region (Bania \etal~2002). Within the context 
of SBBN, D and \3he are completely consistent.

The first challenge to SBBN comes from \4he.  For the SBBN-determined
baryon density the predicted \4he primordial mass fraction is Y$_{\rm P} = 
0.248 \pm 0.001$, to be compared with our adopted value from extragalactic
\hii regions (Olive, Steigman \& Walker 2000) of Y$_{\rm P}^{\rm OSW} = 
0.238 \pm 0.005$.  Agreement is only at the $\sim2\sigma$ level.
Given the unresolved systematic uncertainties in determining \Yp from
the \hii region data, it is not clear at present whether this is a
challenge to SBBN or to our understanding of \hii region recombination
spectra.  As will be seen below, this tension between SBBN D and \4he
can be relieved for nonstandard BBN if the assumption that $S = 1$ 
(\Nnu = 3) is relaxed.

The conflict with the inferred primordial abundance of lithium is 
even more challenging to SBBN.  For \yd = $2.6 \pm 0.4$, [Li] = 
2.65$^{+0.10}_{-0.12}$.  This is to be compared to the estimate 
(see Figure \ref{fig:livsfe}) of [Li] = $2.2 \pm 0.1$ based on a 
sample of  metal-poor, halo stars.  The conflict is even greater 
with the \cite{ryan} estimate of [Li] = 2.09$^{+0.19}_{-0.13}$ 
derived from an especially selected data set.  Unlike the tension
between SBBN and the D and \4he abundances, the conflict between 
D and \7li cannot be resolved by a nonstandard expansion rate (nor, 
by an electron neutrino asymmetry).  Most likely, the resolution 
of this conflict is astrophysical since the metal-poor halo stars 
from which the relic abundance of lithium is inferred have had the 
longest time to mix their surface material with that in their hotter 
interiors, diluting or destroying their prestellar quota of lithium 
(see, \eg \cite{pinsono} and references to related work therein).  

At present SBBN in combination with the limited data set of QSOALS
deuterium abundances yields a $\sim 10\%$ determination of the 
baryon density parameter.  Consistency between the inferred primordial
abundances of D and \3he lends support to the internal consistency 
of SBBN, but the derived primordial abundances of \4he and \7li pose 
some challenges.  For \4he the disagreement is only at the $\sim2\sigma$ 
level and the errors in the observationally inferred value of \Yp are 
dominated by poorly quantified systematics.  However, if the current 
discrepancy is real,  it might be providing a hint at new physics 
beyond the standard model (\eg nonstandard expansion rate and/or 
nonstandard neutrino physics).  Before considering the effects on
BBN of a nonstandard expansion rate ($S \neq 1$; N$_{\nu} \neq 3$),
we will compare the SBBN estimate of the baryon density parameter 
with that from the CBR.

\section{The Baryon Density From The CBR}

\begin{figure}
\centering
\vspace{1.2pc} 
 \epsfysize=3.25truein
  \epsfbox{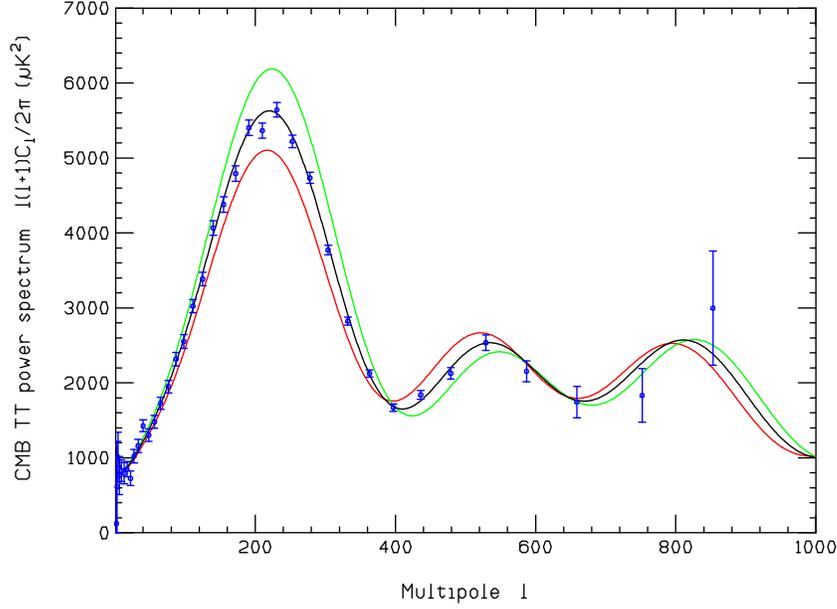}
\caption{The CBR temperature fluctuation anisotropy spectra for three 
choices of the baryon density parameter $\omega_{\rm B} = 0.018$, 0.023, 
0.028, in order of increasing height of the first peak.  Also shown are 
the WMAP data points.
\label{fig:etaspectrum}}  
\end{figure}    

\begin{figure}
\centering
\vspace{1.5pc} 
 \epsfysize=3.25truein
  \epsfbox{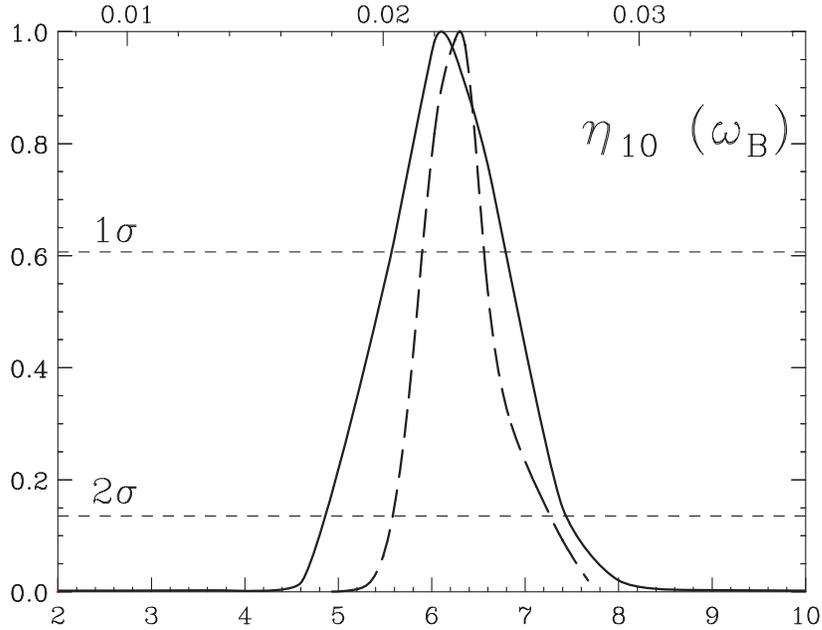}
\caption{The normalized likelihood distributions for the baryon density
parameter $\eta_{10}$ derived from SBBN and the primordial abundance of 
deuterium (solid curve; see \S \ref{sec:bbnbaryometer}) and from the
CBR using WMAP data alone (dashed curve).  The bottom horizontal axis 
is the baryon-to-photon ratio parameter $\eta_{10}$; the top axis is 
the baryon density parameter $\omega_{\rm B}$ = \obh.
\label{fig:etalikbbncmb}}  
\end{figure}    

Some 400 kyr after BBN has ended, when the Universe has expanded and 
cooled sufficiently so that the ionized plasma of protons, alphas, and 
electrons combines to form neutral hydrogen and helium, the CBR photons
are set free to propagate throughout the Universe.  Observations of the
CBR today reveal the anisotropy spectrum of temperature fluctuations 
imprinted at that early epoch.  The so-called acoustic peaks in the
temperature anisotropy spectrum arise from the competition between
the gravitational potential and the pressure gradients.  An increase
in the baryon density increases the inertia of the baryon -- photon
fluid shifting the locations and the relative heights of the acoustic
peaks.  In Figure \ref{fig:etaspectrum} are shown three sets of
temperature anisotropy spectra for cosmological models which differ 
only in the choice of the baryon density parameter $\omega_{\rm B}$.  
Also shown in Figure \ref{fig:etaspectrum} are the WMAP data from 
\cite{wmap}.  It is clear from Figure \ref{fig:etaspectrum} that
the CBR provides a very good baryometer -- independent of that 
from SBBN and primordial deuterium.  Based on the WMAP data alone, 
\cite{barger1} find that the best fit value for the density parameter
is $\eta_{10} = 6.3$ ($\omega_{\rm B} = 0.023$) and that the 
$2\sigma$ range extends from $\eta_{10} = 5.6$ to 7.3 ($0.020 \leq 
\omega_{\rm B} \leq 0.026$).  This is in excellent (essentially 
perfect!) agreement (as it should be) with the CBR-only result 
of \cite{sperg}.  More importantly, as may be seen clearly in 
Figure \ref{fig:etalikbbncmb} (courtesy of D. Marfatia), this 
independent constraint on the baryon density parameter, sampled 
some 400 kyr after BBN, is in excellent agreement with that from 
SBBN (see \S \ref{sec:sbbn}), providing strong support for the 
standard model of cosmology.

The independent determination of the baryon density parameter by
the CBR reinforces the tension between SBBN and the relic abundances
of \4he and \7li inferred from the observational data (see \S 
\ref{sec:sbbn}).  In the context of SBBN, the slightly higher best 
value of $\eta$ from the WMAP data (compared to that from D plus SBBN)
{\it increases} the expected primordial abundances of \4he and \7li
(see Figure \ref{fig:schrplot}), widening the gaps between the SBBN
predictions and the data.  Keeping in mind the observational and
theoretical difficulties in deriving the primordial abundances from
the data, it is nonetheless worthwhile to explore a class of nonstandard
alternatives to the standard model of cosmology in which the early
Universe expansion rate is modified ($S \neq 1$, \Nnu $\neq 3$).

\section{Nonstandard BBN: $S \neq 1$, \Nnu $\neq 3$}\label{sec:nsbbn}

\begin{figure}
\centering
 \epsfysize=4.0truein
  \epsfbox{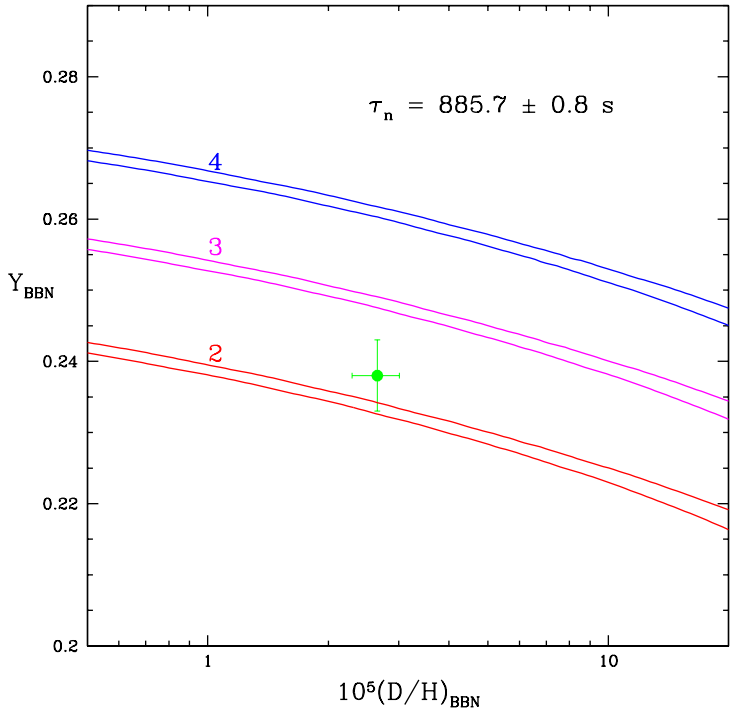}
\caption{The BBN-predicted relation between the \4he mass fraction \Yp 
and the deuterium abundance \yd for three, early-Universe expansion 
rates corresponding to \Nnu = 2, 3, 4.  The filled circle with error 
bars is for the D and \4he primordial abundances adopted here.
\label{fig:hevsd234}}  
\end{figure}    

As outlined in \S \ref{sec:abund}, for fixed $\eta$ as $S$ increases
the BBN-predicted abundances of D, \3he, and \4he increase (less time
to destroy D and \3he, more neutrons available for \4he), while that 
of \7li decreases (less time to produce \7li). Since it is the \4he 
mass fraction that is most sensitive to changes in the early Universe 
expansion rate and, since the SBBN-predicted value of \Yp is too large 
when compared to the data, $S < 1$ (\Nnu $< 3$) is required. For a 
{\it slower} than standard expansion rate the predicted abundances 
of D and \3he {\it decrease} compared to their SBBN values (at fixed 
$\eta$) while that of \7li {\it increases}.  Since the BBN-predicted 
abundance of D increases with decreasing baryon density, a decrease 
in $S$ can be compensated for by a decrease in $\eta$.  For $\eta_{10} 
\approx 6$ and $S - 1 \ll 1$, a good approximation (for fixed D) is 
$\Delta\eta_{10} \approx 6(S - 1)$ (Kneller \& Steigman 2003).  In 
Figure \ref{fig:hevsd234} are shown the \4he -- D (\Yp versus D/H) 
relations for three values of the expansion rate parameterized by 
N$_{\nu}$.  To first order, the combination of $\eta$ and $S$ that 
recovers the SBBN deuterium abundance will leave the \3he abundance 
prediction unchanged as well, preserving its good agreement with the 
observational data.  However, the consequences for \7li are not so 
favorable. The BBN abundance of \7li increases with decreasing $S$ but 
decreases with a smaller $\eta$; the two effects nearly cancel leaving 
essentially the same discrepancy as for SBBN.  For \7li, a nonstandard 
expansion rate cannot relieve the tension between the BBN prediction 
and the observational data.

\begin{figure}
\centering
\vspace{1.2pc} 
 \epsfysize=3.0truein
  \epsfbox{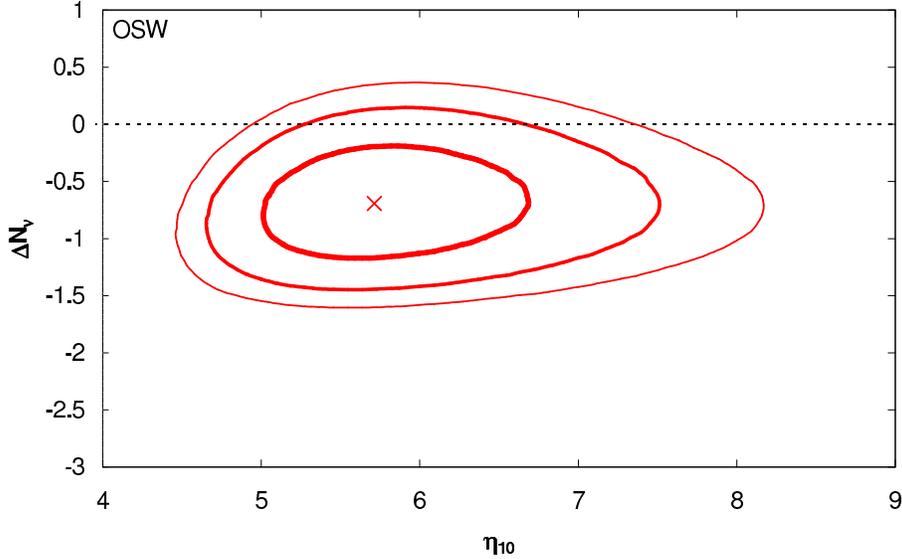}
\caption{The 1$\sigma$, 2$\sigma$, and $3\sigma$ contours in the 
         \Deln -- $\eta_{10}$ plane from BBN and the relic D and 
         \4he abundances.  The best fit values of \Deln and 
         $\eta_{10}$ are marked by the cross.
\label{fig:bbncontour}}  
\end{figure}    

Setting aside \7li, it is of interest to consider the simultaneous
constraints from BBN on the baryon density parameter and the expansion 
rate factor from the abundances of D and \4he; it has already been
noted that for this nonstandard case, D and \3he will remain consistent.
In Figure \ref{fig:bbncontour} are shown the $1\sigma$, $2\sigma$, and 
$3\sigma$ contours in the \Deln -- $\eta$ plane derived from BBN and 
the D and \4he relic abundances.  As expected from the discussion above, 
the best fit value of $\eta$ (the cross in Figure \ref{fig:bbncontour}) 
has shifted downward to $\eta_{10} = 5.7$ ($\omega_{\rm B} = 0.021$).  
While the best fit is for \Deln $= -0.7$ ($S = 0.94$), it should be 
noted that the standard case of \Nnu = 3 is entirely compatible with 
the data at the $\sim2\sigma$ level.

\section{Nonstandard CBR: $S \neq 1$, \Nnu $\neq 3$}\label{sec:nscbr}

\begin{figure}  
\centering
\vspace{1.5pc} 
 \epsfysize=3.5truein
  \epsfbox{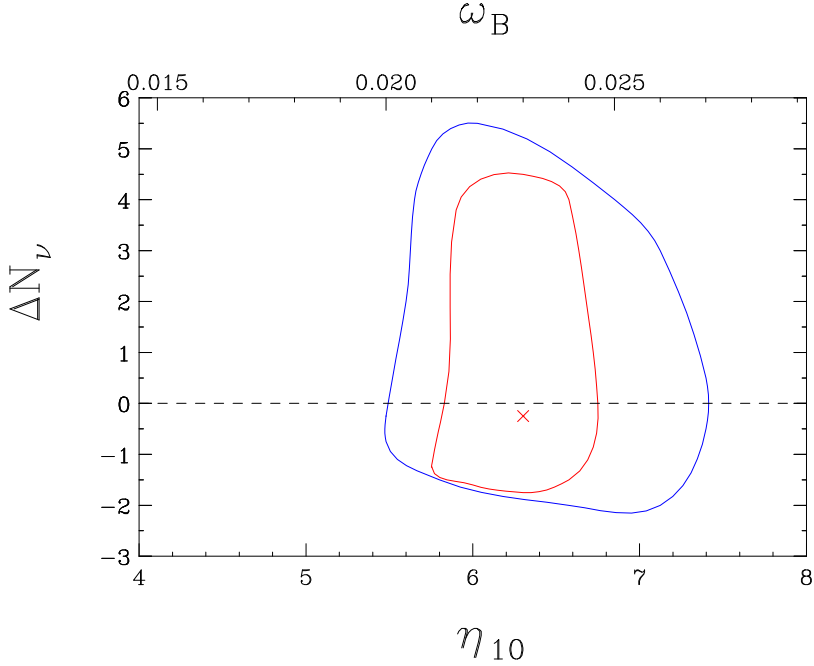}
\caption{The 1$\sigma$ and 2$\sigma$ contours in the $\eta$ --  
         \Deln plane from the CBR ({\it WMAP}) data.  The  best fit 
         point ($\eta_{10} = 6.3$, \Deln $= -0.25$) is indicated by 
         the cross.
\label{fig:cbrcontour}}   
\end{figure}

The CBR temperature fluctuation anisotropy spectrum is sensitive 
to the early-Universe radiation density ($\rho_{\rm R}$) as well 
as to the overall expansion rate.  The early Universe is radiation 
dominated so that $\rho \approx \rho_{\rm R} \propto 1 + 
0.135\Delta$N$_{\nu}$ (see Eq.~\ref{sx} and recall that $\rho 
\propto H^{2}$).  The late Universe is matter dominated (MD) 
($\omega_{\rm M} \equiv \Omega_{\rm M}h^{2}$) and the crossover 
from RD to MD, important for the growth of fluctuations and for 
the age/size of the Universe at recombination, occurs for a redshift
\be
z_{eq} = 2.4 \times 10^{4}\omega_{\rm M}(1 + 0.135\Delta{\rm N}_{\nu})^{-1}.
\label{eq:zeq}
\ee
If the matter content is kept fixed while the radiation content is 
increased, corresponding to a faster than standard expansion rate, 
matter-radiation equality is delayed, modifying the growth of 
fluctuations prior to recombination and, also, the Universe is
younger at recombination and has a smaller sound horizon, shifting 
the angular location of the acoustic peaks.  The degeneracy between 
the radiation density (\Deln or $S$) and $\omega_{\rm M}$ is 
broken by the requirement that $\Omega_{\rm M} + \Omega_{\Lambda} 
= 1$ and the HST Key Project determination of the Hubble parameter 
(see \cite{barger1} for details and further references).  In Figure 
\ref{fig:cbrcontour} are shown the 1$\sigma$ and 2$\sigma$ contours 
in the \Deln -- $\eta$ plane from the CBR (WMAP) data; note the very 
different \Deln scales and ranges in Figures \ref{fig:bbncontour} and 
\ref{fig:cbrcontour}.  As is the case for BBN (see \S \ref{sec:nsbbn}), 
the CBR favors a slightly slower than standard expansion.  However, 
while the ``best'' fit value for the expansion rate factor is at $S < 1$ 
(\Deln $< 0$), the CBR likelihood distribution of \Deln values is very 
shallow and the WMAP data are fully consistent with $S = 1$ (\Deln = 0).

\begin{figure}  
\centering
\vspace{1.5pc} 
 \epsfysize=3.5truein
  \epsfbox{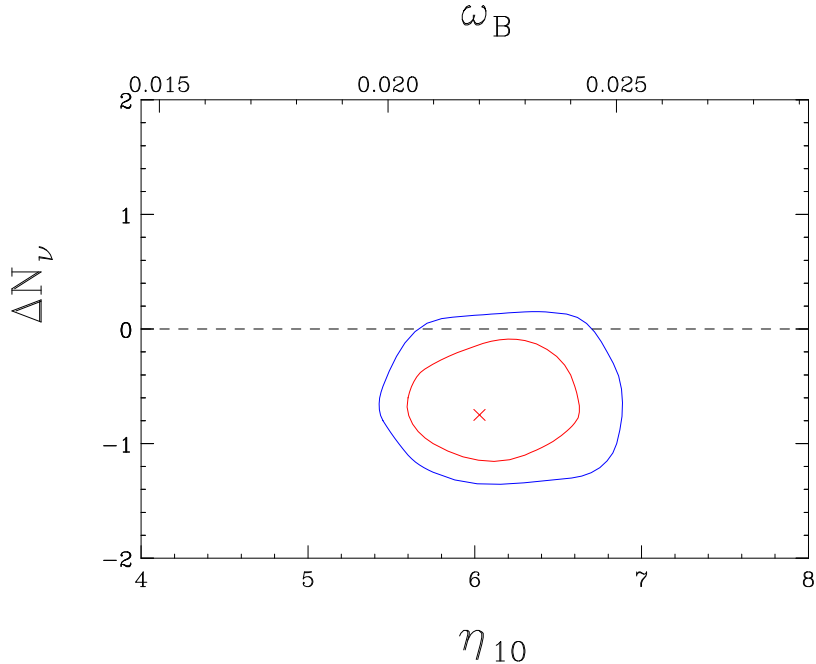}
\caption{As for Figure \ref{fig:cbrcontour}, but for the 
{\it joint} BBN -- CBR fit.  The  best fit point ($\eta_{10} 
= 6.0$, \Deln $= -0.75$) is indicated by the cross.
\label{fig:jointcontours}}   
\end{figure}

Comparing Figures \ref{fig:bbncontour} and \ref{fig:cbrcontour}, 
it is clear that for this variant of the standard cosmology there 
is excellent overlap between the $\eta$ -- \Deln confidence 
contours from BBN and those from the CBR (see Barger \etal~2003a).  
This variant of SBBN ($S \neq 1$) is consistent with the CBR.  In 
Figure \ref{fig:jointcontours} (from Barger \etal~2003a) are shown 
the confidence contours in the $\eta$ -- \Deln plane for a joint 
BBN -- CBR fit.  Again, while the best fit value for \Deln is 
negative (driven largely by the adopted value for Y$_{\rm P}$), 
\Deln = 0 ($S = 1$) is quite acceptable. 

\section{Summary And Conclusions}\label{summary}

As cosmology deals with an abundance of precision data, redundancy 
will be the key to distinguishing systematic errors from evidence 
for new physics.  BBN and the CBR provide complementary probes of 
the Universe at two epochs widely separated from each other and from 
the present. For the standard model assumptions (\Nnu = 3, $S = 1$)
the SBBN-inferred baryon density is in excellent agreement with
that derived from the CBR (with or without the extra constraints
imposed by large scale structure considerations and/or the Lyman
alpha forest).  For this baryon density ($\eta_{10} \approx 6.1$, 
$\omega_{\rm B} \approx 0.022$), the SBBN-predicted abundances of 
D and \3he are in excellent agreement with the observational data. 
For \4he the predicted relic mass fraction is $\sim2\sigma$ higher 
than the primordial abundance inferred from current data, hinting at
either new physics or the presence of unidentified systematic errors.
For \7li too, the SBBN-predicted abundance is high compared to that
derived from very metal-poor stars in the Galaxy.  While the tension
with \4he can be relieved by invoking new physics in the form of a
nonstandard (slower than expected) early-Universe expansion rate, this 
choice will not reconcile the BBN-predicted and observed abundances
of \7li.  When {\it both} the baryon density and the expansion rate 
factor are allowed to be free parameters, BBN (D, \3he, and \4he) 
and the CBR (WMAP) agree at 95\% confidence for $5.5 \leq \eta_{10} 
\leq 6.8$ 
and $1.65 
\leq$~N$_{\nu} \leq 3.03$.  
 
The engine powering the transformation of the study of cosmology from 
its youth to its current maturity has been the wealth of observational 
data accumulated in recent years.  In this data-rich, precision era BBN,
one of the pillars of modern cosmology, continues to play a key role.  
The spectacular agreement between the estimates of the baryon density 
derived from processes  at widely separated epochs has confirmed the 
general assumptions of the standard models of cosmology and of particle 
physics.  The tension with \4he (and with \7li) provides a challenge, 
along with opportunities, to cosmology, to astrophysics, and to particle 
physics.  Whether the resolution of these challenges is observational, 
theoretical or, a combination of both, the future is bright.

\begin{acknowledgments}
I am grateful to all my collaborators and I am happy to thank them 
for their various contributions to the material reviewed here.  Many 
of the quantitative results (and figures) presented here are from 
recent collaborations or discusions with V. Barger, J. P. Kneller, 
H.-S. Lee, J. Linsky, D. Marfatia, K. A. Olive, R. J. Scherrer, V. V. 
Smith, and T. P. Walker.  My research is supported at OSU by the DOE 
through grant DE-FG02-91ER40690. This manuscript was prepared while I 
was visiting the Instituto Astr$\hat{\rm o}$nomico e Geof\' \i sico 
of the Universidade de S$\tilde{\rm a}$o Paulo, and I thank them 
for their hospitality.
\end{acknowledgments}

\end{document}